\documentclass[preprint]{aastex}
\usepackage{natbib}
\slugcomment{Accepted for publication in The Astronomical Journal}
\shorttitle{DETECTION OF EXOPLANETS TRANSITING BROWN DWARFS}
\shortauthors{SENGUPTA}
\received{2016 April 27}
\accepted{2016 July 06}
\begin{document}

\title{POLARIMETRIC DETECTION OF EXOPLANETS TRANSITING T AND L BROWN 
DWARFS}

\author{Sujan Sengupta}
\affil{Indian Institute of Astrophysics, Koramangala 2nd Block,
Bangalore 560 034, India; sujan@iiap.res.in}

\begin{abstract}
While scattering of light by atoms and molecules yields large amount of polarization at
the B-band of both T and L dwarfs, scattering by dust grains in cloudy atmosphere of 
L dwarfs gives rise to significant polarization at the far-optical and  
 infra-red wavelengths where these objects are much  brighter. However, the observable
disk averaged polarization should be zero if the clouds are uniformly distributed and the
object is spherically symmetric.
Therefore, in order to explain the observed large polarization of several L dwarfs, 
 rotation-induced oblateness or horizontally inhomogeneous
cloud distribution in the atmosphere is invoked.
On the other hand, when an extra-solar planet of Earth-size or larger transits
the brown dwarf along the line of sight, the asymmetry induced during the transit 
gives rise to a net non-zero, time dependent polarization. 
Employing atmospheric models for 
 a range of effective temperature and surface gravity
appropriate for T and L dwarfs, I derive the time dependent polarization profiles
of these objects during transit phase and estimate the peak amplitude of polarization
that occurs during the inner contact points of the transit ingress/egress phase.
It is found that peak polarization in the range of 0.2-1.0\%  at I- and J-band 
may arise of cloudy L dwarfs occulted by Earth-size or larger exoplanets. 
Such an amount of polarization is higher than that can be produced by rotation-induced
oblateness of even the rapidly rotating L dwarfs. Hence, I suggest that time
resolved imaging polarization should be a potential technique to detect transiting exoplanets
around L dwarfs.
    
\end{abstract}


\section{INTRODUCTION}

 Twenty years after the first confirmed discovery of
planets outside the solar system \citep{wolszczan92,wolszczan94,mayor95},
about 3000 planets orbiting around stars of different spectral types are
detected. Out of a wide variety of planets with different mass, size and
surface temperature, the recently discovered small and possibly rocky
planets that may have surface temperature appropriate for water to exist
in liquid state have become the objects of utmost interest. 
 They have compelled to change the focus into
detecting planets that may have favorable environment to
harbor life.

  Over 80\% of the stellar population in the Galaxy consists of red  dwarfs
 and brown dwarfs \citep{apai13}. About 15\%  of
the objects in the solar neighborhood are brown dwarfs \citep{bihain16}.
Statistical analysis of data obtained by using the Kepler space telescope indicates
that the occurrence rate of small and potentially rocky planets with radius ranging
from 0.5 to 2.0 $R_{\oplus}$ ($R_{\oplus}$ is the radius of the
Earth) around M dwarf stars is about 0.51 per star
\citep{dressing13, koppa13}.
A good fraction of these rocky planets are believed to be located 
within the habitable zone of their star. Therefore, a large
number of habitable planets are expected in the solar neighborhood  and
in our galaxy. However, unlike a solar type of star, M dwarfs are very active
as they become fully convective from mid to late spectral type. They also have
intense magnetic field ranging between 3-9 kG \citep{reid05}. As a consequence frequent flare
and emission of intense x-ray and extreme ultraviolet ray are very common. Therefore,
the atmosphere of a planet in the habitable zone of M dwarf star is unlikely to have
favorable condition for life to survive \citep{lissa07,sengupta16a, mohanty16}. Such a situation
strongly motivate the search for habitable planets around brown dwarfs.

Brown dwarfs are believed to be born like a star but fail to become a 
main-sequence star because they have masses sufficient to ignite deuterium burning
but insufficient to enter into the hydrogen burning main sequence. Therefore,
they inhabit the realm between the least massive stars and the most
massive planets. The source of the observed radiation is the gravitational
potential energy during the formation and contraction phase. With the
decrease in temperature of an individual
brown dwarf, it progressively passes through spectral types
ranging from late M through the L, T and Y sequences. Detailed
reviews on the properties of brown dwarfs are given by \cite{chabrier00,basri00,burrows01,kirkpatrick05,
luhman12,marley15}. Synthetic spectra of cloud-free model atmospheres fit well 
the observed spectra  of field T dwarfs later than about type T4
because dust grains condensation takes place well bellow the
photosphere and therefore are not an important opacity source 
\citep{stephens09}. On the other hand, inclusion of condensate cloud of various 
species such as iron, forsterite ($\rm Mg_2SiO_4$), and
enstatite ($\rm MgSiO_3$) explains the spectra of L dwarfs \citep{cushing08,marley15}.

Similar to the case of the Sun and cool stars, thermal radiation of both T and L brown dwarfs
should be linearly polarized at the near-optical wavelengths, e.g., in the 
B-band due to Rayleigh scattering by atoms and molecules. On the other hand, the
presence of dust grains in the visible atmosphere of L dwarfs gives rise to significant amount
of linear polarization in the far-optical and in the infra-red regions. However,
the net observable polarization would be zero if the object is spherically symmetric
and the clouds are distributed uniformly. Linear polarization has been detected in the
optical bands from a good number of L dwarfs covering almost the entire range of spectral
types L0--L8 \citep{men02,oso05,tata09,oso11,miles13}. Scattering by horizontally
homogeneous clouds in the atmosphere of a rotation-induced oblate L dwarf can explain the
observed polarization 
\citep{sengupta01,sengupta03,sengupta05,sengupta10}. Imaging polarimetric
 data of L dwarfs shows increase in the amount of polarization with the increase
in spin rotation velocity \citep{miles13} implying that the asymmetry due to rotation-induced
oblateness plays an important role in the linear polarization of fast rotating  dusty L dwarfs.
However, in agreement with the analysis of
\cite{sengupta09}, no polarization in far-optical and near infra-red is detected
from any T dwarf to date.     

Apart from the asymmetry that may arise due to rotation-induced 
oblateness of the stellar disk or/and horizontally inhomogeneous 
clouds in the atmosphere, asymmetry in the stellar disk can also be produced
by a transiting planet that blocks the stellar disk partially. Consequently,
transit of planet can also give rise to net non-zero disk integrated 
polarization. Such transit or occultation polarization of stars of 
different spectral types has been discussed in detail by a few  authors  
 \citep{car05,wik14,kostogryz11}. 

Giant exoplanets around brown dwarfs are discovered by direct 
imaging \citep{Chau04,todorov10,gauza15,stone16}, radial velocity method \citep{joergens07} and
by gravitational microlensing method \citep{han13}. Recently a Venus-size planet
has been discovered by microlensing \citep{udalsky15}. These discoveries clearly
imply that formation of exoplanets with size ranging from  Jovian to
sub-Earth is possible around brown dwarfs either through binary star formation mechanism or through
the scale-down core-accretion mechanism of planet formation around a star. Therefore,
a large number of exoplanets around brown dwarfs are awaiting to be detected \citep{apai13}.
Discovery of 
these planets may highly populate the number of exoplanets.
In fact, such a realization prompted to propose strategies for detecting habitable
planets around brown dwarfs \citep{caba02,caba10,belu13}. However, as pointed out by
\cite{udalsky15}, owing to the limitation of available technology, substantially 
low mass planetary companions to brown dwarfs can only be discovered at present
by using the gravitational microlensing technique. Obviously, the main constrain 
in detecting Earth-size planet around brown dwarfs through transit or radial velocity
method is the extreme faintness of these objects.

   In this paper, I show that time resolved image polarimetry can be a potential
tool to detect planets transiting cloudy L dwarfs. Using detailed atmospheric models
for cloudy L dwarfs, the scattering polarization is calculated and it is shown that the asymmetry
induced by a transiting Earth-size planet gives rise to significant amount of disk-integrated
linear polarization in both I- and J-bands at the inner contact points of transit
ingress/egress phase that may be detected by the existing imaging polarimeters.     

   In the next section, I briefly describe the formalism adopted to estimate 
the time-resolved transit polarization profile of T and L dwarfs. In section~3,
I discuss the results following by our conclusions in section~4.

\section{METHOD FOR CALCULATING THE TRANSIT POLARIZATION}

The net polarization of the L and T dwarfs during the transit is calculated by multiplying
the intensity and scattering polarization at each radial point along the disk of the object
with  the fractional circumference occulted by
the projection of the planet over the surface of the primary dwarf.  
The disk-averaged  polarization during the transit phase or the transit polarization is given by
\cite{sengupta16,wik14,car05}. Here we use the formalism  presented in \cite{sengupta16}.
Accordingly,
\begin{eqnarray}
p_(t) =\frac{1}{F}\int^{r_2}_{r_1}2\sqrt{\frac{[(1-\mu^2)^{1/2}-r_m(t)]^2-w^2}{1-\mu^2}}I(\mu)
p(\mu)\mu d\mu,
\end{eqnarray}
where  $t$ is the time since mid-transit,
 $F$ is the flux of the unobscured  dwarf, $I(\mu)$ and
$p(\mu)$ are the specific intensity and polarization respectively along $\mu$,
$\mu=\cos\theta=\sqrt{1-r^2}$ with $\theta$ being 
the angle between the normal to the surface of the primary and the line of sight and
$r$ being the radial points along the disk of the primary, $0\leq r \leq 1$,     
$r_1=\sqrt{1-[r_m(t)+w]^2}$ and $r_2=\sqrt{1-[r_m(t)-w]^2}$,
$r_m(t)$ is the instantaneous position of the center of the planet and is given by
\begin{eqnarray}
r_m(t)=\left[b^2+4\left\{(1+w)^2-b^2\right\}\left(\frac{t}{\tau}\right)^2
\right]^{1/2}.
\end{eqnarray}
In the above expression, the impact parameter $b$ for a circular orbit of radius $a$ is given by
$b=a\cos i/R_\star$, where $i$ is the orbital inclination angle of the planet
and $R_\star$ is the radius of the L or the T dwarf, $w=R_P/R_\star$ is the ratio of the planetary 
radius ($R_P$) to the radius of the primary.
The transit duration $\tau$ is given by \cite{scharf09}
\begin{eqnarray}
\tau=\frac{P}{\pi}\sin^{-1}\left[\frac{R_\star}{a}\left\{\frac{(1+w)^2-b^2}{1-
\cos^2 i}\right\}^{1/2}\right]
\end{eqnarray}
Transit of planet can occur only if the inclination angle 
$i\geq \cos^{-1}\left(\frac{R_\star+R_P}{a}\right)$.  

In order to calculate the specific intensity $I(\mu)$ and scattering polarization $p(\mu)$,
I have employed one-dimensional, non-grey, hydrostatic and radiative-convective
atmospheric models 
for a range of effective temperature $T_{\rm eff}$ and surface gravities $g$ appropriate for 
T and L dwarfs \citep{ack01,marley02,freed08,Sau08}.  
For T dwarfs, cloudless model atmosphere generally
reproduce the spectra of most T dwarfs with $T_{\rm eff} < 1200-1300\,\rm K$ \citep{stephens09}.
 In the present work, I consider cloudless
models for T dwarfs appropriate for spectral types later 
than about T3 \citep{stephens09}. However,  
condensates are included in the chemical equilibrium  calculation
\citep{freed08}. For T dwarfs, I choose models with $T_{\rm eff}$ = 1300, 1100, 900,
and 700 K for a fixed value of g=1000 m$s^{-2}$. 

For the relatively hotter L dwarfs, spatially uniform dust cloud is included.
The efficiency of sedimentation of cloud particles in the atmospheric models
is controlled through a scaling factor $f_{\rm sed}$.
In the present work I adopt $f_{\rm sed}=2.0$ \citep{cushing08,stephens09}. 
For L dwarfs, I choose models with $T_{\rm eff} = 1400$, 1600, 1800, 2000 and
2200 K for a fixed surface gravity g=1000 m$s^{-2}$. A few representative cases
for g=300 m$s^{-2}$ are also presented for comparison. 
The atmosphere models employed here fits reasonably well
the spectra and photometry of a large number of T and L  dwarfs at a wide range of
wavelengths covering near optical to mid-infrared regions \citep{marley15}.

 The gas and dust opacity, the temperature-pressure profile
and the dust scattering asymmetry function
averaged over each atmospheric pressure level are computed by the atmospheric code.
In order to calculate the two non-zero Stokes parameters I and Q  in a locally
plane-parallel medium , these input data are used in a multiple
scattering polarization code that solves the radiative transfer equations in vector form.
It is worth mentioning that
for linear polarization of an axially symmetric radiation field, the other two Stokes
parameters, U and V are zero. 
For the cloudy L dwarfs, the angular distribution of the photons before and after
 scattering is calculated by using a combined Henyey-Greenstein-Rayleigh phase
matrix \citep{Liu06} while for  cloudless T dwarfs, Rayleigh phase matrix is sufficient
to describe the scattering by atoms and molecules \citep{chandra}. The detailed formalisms as
well as the numerical methods for calculating the angle dependent total
and polarized intensity $I$ and $Q$  are described  in \cite{sengupta09}.
In fact, in order to calculate the transit or occultation polarization profiles presented
here, I have used the same values of $I(\mu)$ and $p(\mu)$ that are used in
\cite{sengupta09} for T dwarfs and in \cite{sengupta10} for L dwarfs.

  If the reflected light of the planet is polarized due to the presence of cloud in 
the planetary atmosphere, a small amount of phase dependent polarization may arise
at far optical and infrared wavelenghts.
However, because of very small planet-to-dwarf flux ratio, the amount of such 
polarization should be insignificant as compared to
the polarization of the brown dwarf \citep{seager00, sengupta06, bott16}. Hence, we ignore any contribution by the
planet to the net polarization of the system.
   
 Although the atmospheric code derives different radius of the brown dwarf for different
surface gravity and effective temperature, in order to derive the transit duration of a
planet with a fixed orbital separation $a=0.01$ AU, I assume that the radius
of the brown dwarf to be 1$R_J$ where $R_J$ is the radius of Jupiter. However, for
a given surface gravity, this does not alter the transit duration significantly.
For example, with a fixed surface gravity ${\rm g}=1000 {\rm ms^{-2}}$,
the transit duration of an Earth-size planet orbiting a brown dwarf with radius 1.5$R_J$ is 
0.72 hrs when the orbital inclination angle is $90^o$ while the transit 
duration of an Earth-size planet orbiting the primary with radius 1$R_J$ is 0.74 hrs. 
It must be emphasized here that this is just a representative case. 
The amount of polarization originated due to transit is not affected by the orbital
distance or period of the planet. The orbital separation and the orbital period
determines the transit duration which provides the interval between
the two successive amplitudes of polarization that occur at the inner contact points of
the transit ingress/egress phase as depicted in Figure~1. The amount of transit polarization
depends on the ratio between the radii of the planet and the 
L or T  dwarf.  

\section{RESULTS AND DISCUSSIONS}

The center-to-limb variation in the polarization across any stellar disk
arises due to the scattering albedo which is determined by the
contribution of scattering opacity
to the total opacity in the atmosphere \citep{harrington69}. 
 The polarization is zero at the center ($\mu=1$) of the stellar disk
and is maximum at the limb ($\mu=0$). Linear polarization in cool stars
arises by scattering of light with atoms and molecules and the maximum amount of 
polarization that occurs at the stellar limb is usually very small.
 Polarization for solar type stars varies 
from a few times of $10^{-4}$ near the limb to a few times of $10^{-6}$ near 
the center at near-optical (B-band) wavelengths \citep{fluri99}. At longer
wavelengths, the polarization is extremely small. 
The scattering polarization of cloudless T dwarfs does not
differ significantly from that of a cool star but large amount of
polarization in longer wavelengths arises due to dust scattering in cloudy L dwarfs.

As mentioned earlier, in the absence of clouds in
the visible atmosphere, the polarization of T dwarfs arises due to Rayleigh scattering
by atoms and molecules. Therefore the largest amount of polarization arises at shorter
wavelengths. Figure~2 presents the B-band transit or occultation polarization of T dwarfs 
with different effective temperatures. For a fixed surface gravity, the gas density increases  
with the decrease in $T_{\rm eff}$ and hence the scattering probability increases 
yielding into higher polarization.  It is worth mentioning that 
the amount of scattering polarization at different angular points is
determined by the atmospheric models invoked.

The double peaked polarization profile, a general feature of transit polarization of any  object,
arises because of the fact that
the maximum polarization occurs near the inner contact points of transit ingress/egress phases. For
central transit i.e., when the inclination angle $i=90^o$, the projected position of 
the center of the planet coincides with the center of the star during mid transit. This
gives rise to a symmetry to the projected stellar disk. Hence
the disk integrated polarization for central transit is zero during mid transit. 
 Owing to increase in asymmetry to the stellar disk, the
polarization increases as the planet moves from the center ($t=0$) to the limb (t=$\pm\tau/2$)
of the stellar disk. However, when the eclipse is off center, i.e., when $i\leq 90^o$,
the polarization is non-zero during the whole transit epoch including the mid transit time.
Nevertheless, the peak polarization is independent of the orbital inclination angle
but depends on the ratio of the planetary to stellar radii.

 As presented in Figure~2, for an Earth-size transiting planet, the B-band peak polarization
at the inner contact points of transit ingress/egress phase is 0.06\%  for T dwarfs with 
$T_{\rm eff}=700$K.  The peak polarization
becomes about 0.12\% when the size of the transiting planet is three times the size of
the Earth. The asymmetry in the stellar disk increases with the increase in the ratio between the
planetary and stellar radii and hence the disk-integrated polarization increases.  
 The transit duration
depends on the size of both the primary  and the planet, on the orbital distance of the planet 
from the star or brown dwarf, on the orbital period and on the planetary orbital inclination angle. 
Therefore, for a given orbital distance, the peak polarization occurs at different 
time for different size ratio and inclination angle.

   The amount of transit polarization of T dwarf is maximum at B-band and the peak polarization is
very small, between 0.02 to 0.06 \% for an Earth-size transiting planets. Since the
object is also faint at near-optical, 
detecting such low polarization could be challenging. It's worth mentioning 
here that B-band polarimetric
observation of brown dwarfs is not reported till date and no polarization is detected for any
T dwarf at longer wavelengths \citep{jensen16}. 

 On the other hand, formation of dust grains in the cloudy atmosphere of L dwarfs 
provides an additional scattering opacity to the gas opacity.
 The thermal radiation of the cloudy L dwarfs is polarized by dust 
scattering giving rise to large value of limb polarization in the far-optical and infra-red
wavelengths. Figure~3 presents the J-band transit polarization of L dwarfs. Although the 
overall feature of the transit polarization profile remains the same to that of T dwarfs
at  B-band, the degree of polarization increases by several times at J-band.
As demonstrated by \cite{sengupta10,marley11}, the degree of
polarization of cloudy L dwarfs and self-luminous exoplanets  depends on the 
effective temperature and  the surface gravity of the objects.
The scattering opacity is determined by a balance
between the downward transport by sedimentation and upward turbulent diffusion of
condensates and gas and hence the polarization varies with different
effective temperature and surface gravity.  
 Figure~3 shows that L dwarfs
of mid-spectral type corresponding to $T_{\rm eff}=1600-1800$K produce the largest amount
of polarization at the inner contact points of transit ingress/egress phase when the surface gravity
g=1000 m$s^{-2}$. Polarization reduces with the decrease in surface gravity and effective 
temperature. At the same time, for a given surface gravity polarization reduces when the
effective temperature of the object rises above 1800 K.   
An increase in temperature causes the cloud base to shift upward yielding a smaller column
of dust grains in the observed atmosphere and hence the polarization decreases with the
increase in effective temperature. 

    Although the scattering polarization is sensitive to the effective temperature 
and the surface gravity, the amount of transit polarization strongly depends on
the asymmetry in the stellar disk produced during the transit phase. This asymmetry 
is governed by the size of the transiting planet as
compared to the size of the L dwarf. 
Figure~4 presents the transit polarization at I- and J-bands of a L dwarf
with fixed surface gravity and effective temperature but for different sizes of the
transiting planet. As the asymmetry increases with the increase in the size of the
transiting planet, the amount of polarization at the inner contact points of the ingress/egrees
phase increases linearly. Figure~4 also indicates
that the amount of polarization in J-band is higher than that in I-band. In fact, the
peak amplitude of polarization in J-band is almost double to that in I-band.        
The adopted cloud model, e.g., the size distribution, the number density of the dust grains
as well as the location of the cloud base and deck dictates the amount of polarization at
different wavelengths.

   As mentioned before, the orbital inclination angle of the transiting planet does not determine
the amount of polarization at the inner contact points of the transit ingress/egress phase but affects
the time interval between the two successive peaks of polarization. Figure~5 shows the J-band
transit polarization due to a transiting planet having different orbital inclination angles.
As the inclination angle decreases from its maximum value of $90^o$, the transit path of 
the planet shifts towards the edge of the primary shortening the transit period. The amount
of polarization at the mid transit (t=0) increases with the decrease in the inclination angle.
As the distance between the inner contact points of the ingress and the egress phase shortens,
the time interval between the occurrence of the two polarization peaks reduces
and subsequently the two peaks merge into one central peak when the inclination angle attains
its minimum value beyond which a full transit is not possible. This is consistent with the
results presented by \cite{car05} for solar type of stars. A partial transit however, reduces
the degree of disk integrated polarization. 

Brown dwarfs are fast rotators and rotation causes departure from  sphericity. The asymmetry
induced by rotation can yield into net non-zero disk-integrated polarization \citep{sengupta01}. 
However, as shown by \cite{sengupta10}, the net degree of linear polarization in J-band of a
brown dwarf with surface gravity g=1000 m$s^{-2}$ yields as little as 0.06\% of polarization
even if the projected rotational velocity is about 50 km/s. In order to produce 0.2 \%  of
polarization, a brown dwarf needs to have an oblateness greater than 0.2. For a surface gravity
of 1000 ${\rm ms^{-2}}$, this can never be achieved as the rotation velocity cannot exceed the 
breaking velocity.       
Therefore the amount of polarization at the inner contact points of transit ingress/egress phase
 clearly implies that the asymmetry produced by an Earth-size transiting planet is 
higher than
 that caused by rotation-induced oblateness of even a fast rotating brown dwarfs.
Although, rotation-induced oblateness 
explains the observed I-band polarization of many L dwarfs, the required asymmetry
needs a surface gravity much lower than 1000 m$s^{-2}$ \citep{sengupta10}. 
The synthetic spectra of brown dwarfs poorly
constrain the surface gravity - within a wide range of
300-3000 ${\rm ms^{-2}}$. But some L dwarfs show such a large
polarization that it can be explained only if the spin rotation velocity is extremely high
even if the surface gravity is taken to be 300 ${\rm ms^{-2}}$. 
For example, the observed degree of polarization 2MASS J2158-1550 (L4.0) and
2MASS J1807+5015 (L1.5) are 1.38 \% and 0.7\%  respectively and so a spin rotation
velocity of more than 100 ${\rm kms^{-1}}$ is required to explain such high polarization.
 None of these L dwarfs showed optical variability  and hence rules out 
high inhomogeneity in atmospheric cloud distribution.
On the other hand, I-band polarization as high as 2.45\%  was observed from  
2MASS J2244+2043 (L6.5).  This object is found to be variable at 4.5 micron \citep{morales06}.
However, it is unlikely that cloud inhomogeneity can account for such a large amount of
polarization.  As pointed out by \cite{sengupta10}, quite a few L dwarfs that are optically 
variable do not show polarization implying cloud inhomogeneity that may be responsible for
variability cannot contribute significantly to yield detectable polarization. 
Transit by a Neptunian planet may
explain such high polarization and time resolved imaging polarimetry may verify such a
possibility. If the high polarization is caused by rotation-induced oblateness, the
amount of polarization should remain time-independent. On the other hand, cloud
inhomogeneity should cause variable polarization. In both the cases, the contribution
due to rotation-induced oblateness or cloud inhomogeneity may be estimated when
the primary source is out of the transit phase.     
   
\section{CONCLUSIONS}

Assuming the transit of an exoplanet around  cloudy L dwarfs and cloudless T dwarfs, 
I have presented the transit polarization profiles of brown dwarfs
and estimated the peak polarization at the inner contact points of transit ingress/egress phase
for a large number of model atmospheres and planetary radii. For this purpose, the atmospheric
 models that fit the observed spectra of T and L dwarfs are employed. The B-band linear 
polarization due to scattering by atoms and molecules in the atmosphere of T dwarfs
and I- and J-bands linear polarization due to scattering by dust grains in the atmosphere of L dwarfs
are calculated numerically. The present investigation
implies that significant amount of transit polarization may arise in both I- and J-bands of
a cloudy L dwarf if an Earth-size or larger planet transits it giving rise to asymmetry in
the stellar disk. However, the transit polarization
of the cloudless T dwarfs is significant only in the B-band where these objects are  extremely faint and hence
detection of polarization signal is difficult.
Therefore, it is suggested that time resolved imaging polarimetry should be a potential
technique to detect small rocky planets around L dwarfs. 

  Our model estimations imply that for L dwarfs with spectral types ranging from L3 to L7,
the peak polarization at the inner contact points of the transit ingress/egress phase should be
0.5-1.0\% in J-band and 0.2-0.5\% in the I-band for transiting planets of 
size $1-3R_\oplus$. Such an amount of polarization can easily be detected by several
existing facilities including FORS1 onboard VLT and LIRIS onboard WHT.  

\section{Acknowledgements} 
 
I thank the reviewer for a critical reading of the manuscript and for providing many useful suggestions.

\clearpage
\begin{figure}
\includegraphics[angle=0.0,scale=0.8]{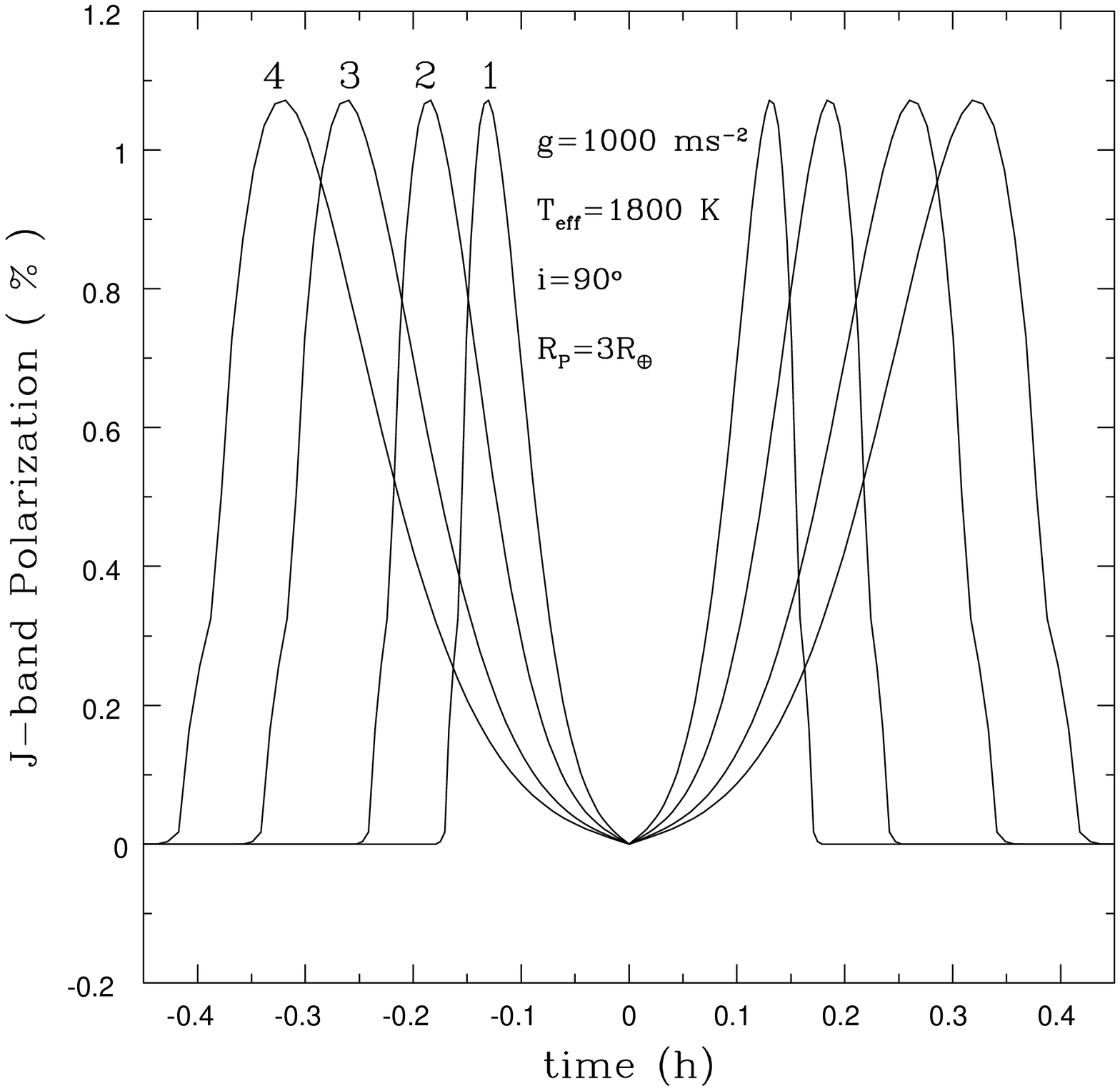}
\caption{Transit polarization of L dwarf by planet with different orbital
separations $a$. The labels represent the polarization profile for
different values of $a$ in AU.  1. $a=0.005$, 2. $a=0.01$, 3. $a=0.1$ and
4. $a=0.2$. A Jupiter size L dwarf is assumed. 
\label{fig1}}
\end{figure}

\begin{figure}
\includegraphics[angle=0.0,scale=0.8]{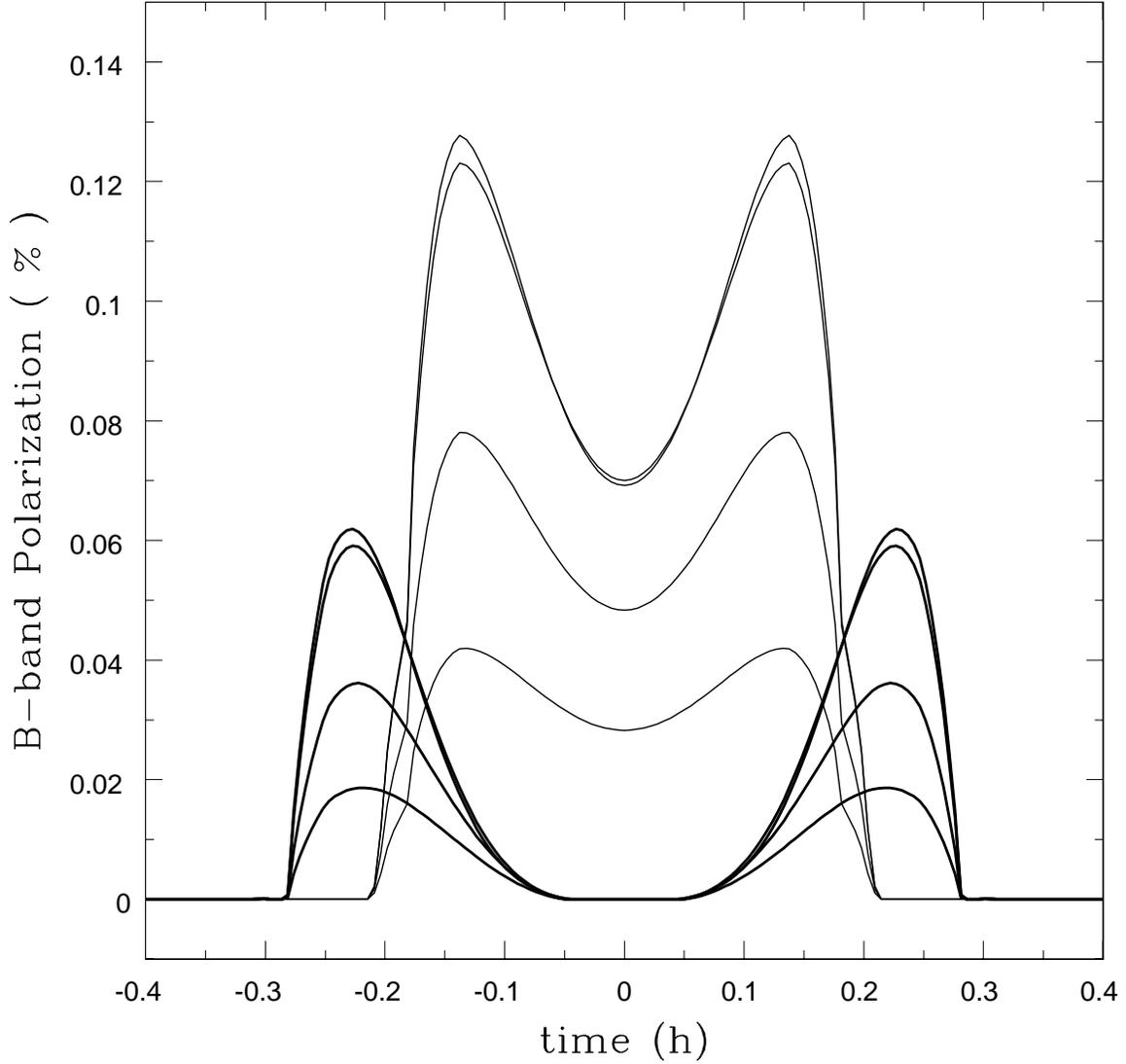}
\caption{Disk integrated B-band polarization of T dwarfs during the
transit of an exoplanet. From top to bottom,  the thick and thin solid
lines represent T dwarf models with $T_{eff}=$700, 900, 1100 and 1300 K 
respectively. The thick solid lines represent
models for an Earth-size transiting planet with $90^o$  orbital inclination
angle and the thin solid lines represent models for
transiting exoplanet with radius $3R_\oplus$ and inclination angle $89^o$.
For all cases, the radius of the T dwarf is set to $1R_J$ and g=1000
m$s^{-2}$. The orbital distance of the exoplanet is 0.01 AU.
\label{fig2}}
\end{figure}

\begin{figure}
\includegraphics[angle=0.0,scale=0.8]{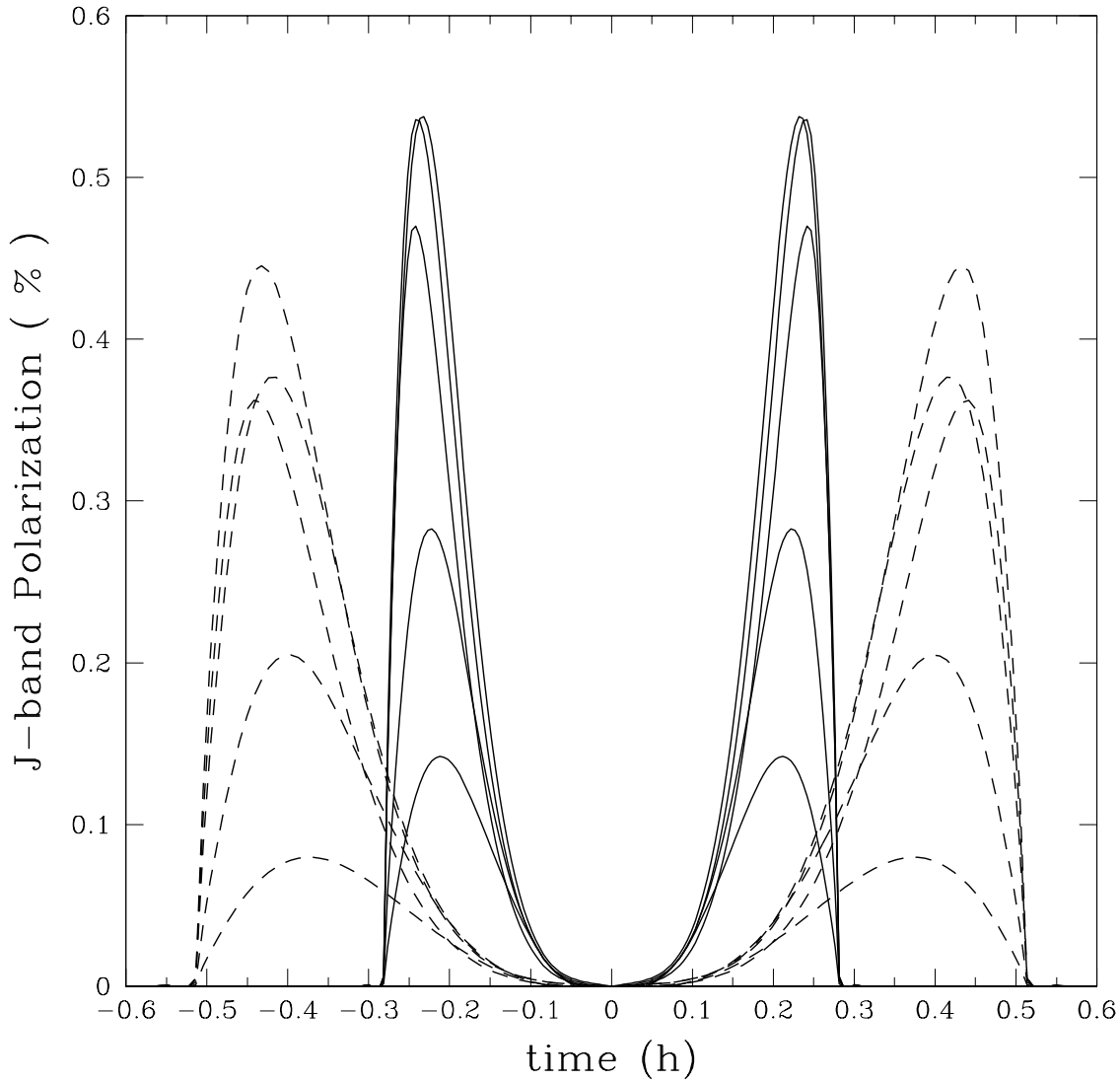}
\caption{Disk integrated J-band polarization of a cloudy L dwarfs during the
transit of an Earth-size planet with orbital inclination angle $i=90^o$ and
orbital distance $a=0.01$ AU.
Solid lines represent L dwarf with surface gravity g=1000 m$s^{-2}$ and
dashed lines represent L dwarf with g=300 m$s^{-2}$. From top to bottom the
lines represent L dwarfs with $T_{eff}=$1600, 1800, 1400, 2000 and 2200 K
respectively.
\label{fig3}}
\end{figure}

\begin{figure}
\includegraphics[angle=0.0,scale=0.8]{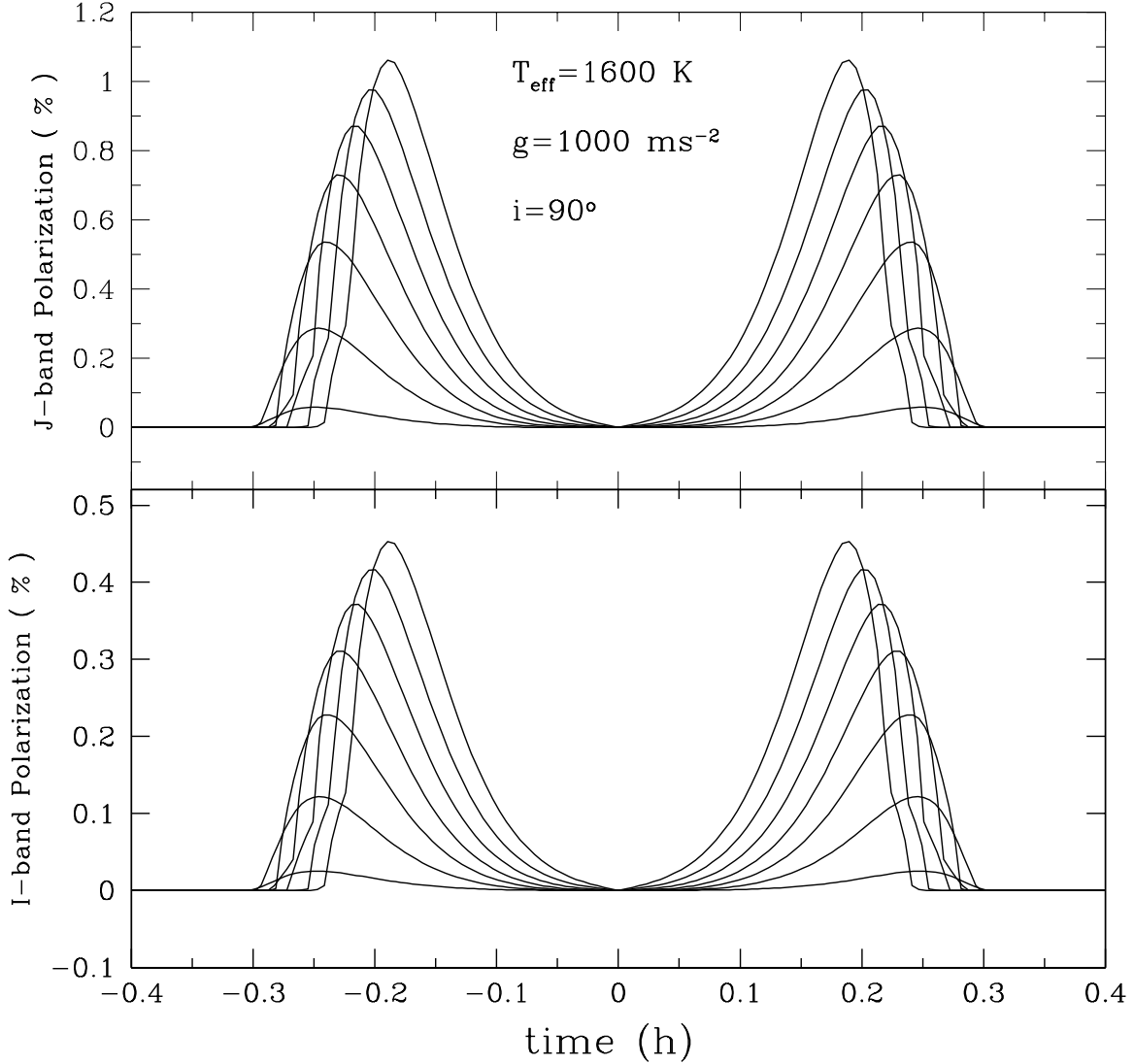}
\caption{ Disk integrated J- and I-band polarization of a L dwarf with
$T_{eff}=1600$K and g=1000 m$s^{-2}$ during the transit of exoplanets
with different sizes but with a fixed orbital inclination angle $i=90^o$
and orbital distance $a=0.01$ AU. In both the panels, the solid lines from
top to bottom  represent polarization by transiting exoplanet of radius 0.1,
0.5, 1.0, 1.5, 2.0, 2.5 and 3.0 $R_\oplus$ respectively
where $R_\oplus$ is the radius of the Earth.
\label{fig4}}
\end{figure}

\begin{figure}
\includegraphics[angle=0.0,scale=0.8]{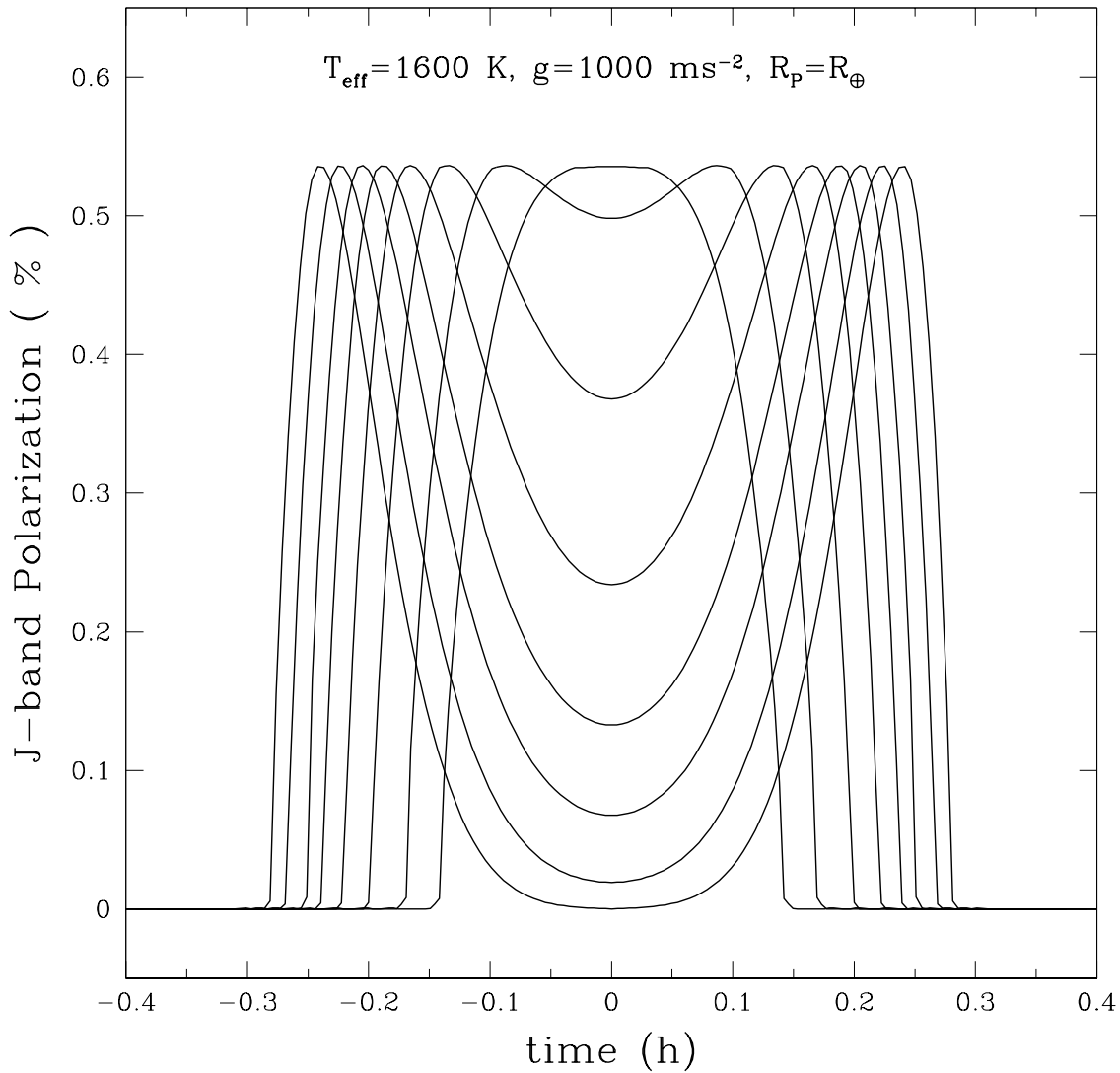}
\caption{ Disk integrated J-band polarization of a L dwarf with
$T_{eff}=1600$K and g=1000 m$s^{-2}$ during the transit of an Earth-size planet
orbiting at a distance of 0.01 AU but with different orbital inclination
angles. From bottom to top at 0 h, the solid lines represent models with
i=90.0, 89.3, 89.0, 88.8, 88.6, 88.4, 88.2 and $88.0^o$ respectively.
\label{fig5}}
\end{figure}

\end{document}